\documentclass[twocolumn,showpacs,preprintnumbers,amsmath,amssymb,nofootinbib]{revtex4}
\usepackage{graphics}
\usepackage[varg]{txfonts} 
\usepackage[latin1]{inputenc}
\begin{document}
\title{Nonextensive critical effects in relativistic nuclear
mean field models}
\author{Jacek Ro\.zynek} \email{rozynek@fuw.edu.pl}
 \author{Grzegorz Wilk} \email{wilk@fuw.edu.pl}
\affiliation{The Andrzej So\l tan Institute for Nuclear Studies,
Ho\.za 69, 00681 Warsaw, Poland}
\begin{abstract}We present a possible extension of the usual
relativistic nuclear mean field models widely used to describe
nuclear matter towards accounting for the influence of possible
intrinsic fluctuations caused by the environment. Rather than
individually identifying their particular causes we concentrate on
the fact that such effects can be summarily incorporated in the
changing of the statistical background used, from the usual
(extensive) Boltzman-Gibbs one to the nonextensive taken in the
form proposed by Tsallis with a dimensionless nonextensivity
parameter $q$ responsible for the above mentioned effects (for $q
\rightarrow 1$ one recovers the usual BG case). We illustrate this
proposition on the example of the QCD-based Nambu - Jona-Lasinio
(NJL) model of a many-body field theory describing the behavior of
strongly interacting matter presenting its nonextensive version.
We check the sensitivity of the usual NJL model to a departure
from the BG scenario expressed by the value of $| q - 1|$, in
particular in the vicinity of critical points.
\end{abstract}
\maketitle
\section{Introduction}
\label{intro}

In all studies of relativistic properties of nuclear matter, mean
field models are usually the models of first choice \cite{SW,NJL}.
These models use a statistical approach based on Boltzmann-Gibbs
(BG) statistics which is, strictly speaking, only correct when the
corresponding heat bath is homogeneous and infinite. These
conditions are by no means met in realistic situations in which
nuclear matter occurs. Usually one encounters some inherent
problems arising, for example, from the smallness of the
collisional systems and their rapid evolution. These, among other
things, render the spatial configuration of the system far from
uniform and prevent global equilibrium from being established (cf.
\cite{departure} and references therein). As a result, some
quantities become non extensive and develop power-law tailed
rather than exponential distributions. The widely used way to
account for these effects is to resort to a nonextensive
statistics, known as $q$-statistics \cite{T}. The new
phenomenological nonextensivity parameter $q$ occurring there is
supposed to account for all possible dynamical factors violating
the assumptions of the usual BG statistics. This is recovered in
the limit of $q \rightarrow 1 $. Because it enters into the
respective formulas of the particular dynamical model used for a
given investigation, it allows for a simple phenomenological check
of the stability of the model against possible deviations from the
BG approach.

So far, applications of the nonextensive approach are numerous and
cover all branches of physics \cite{applications}. These include
high energy multiparticle production processes (cf.,
\cite{multipart}) and different aspects of nuclear and quark
matter \cite{AL,LPQ,BiroQ}. The nonextensive framework can also be
derived from a special treatment of kinetic theory investigating
complex systems in their nonequilibrium stationary states
\cite{qBiro}. Some examples of more specialized topics can be
found in \cite{todos} and references therein. For an illustration,
the Tsallis distribution, $h_q(E)$, and BG distribution, $f(E)$,
are connected as follows:
\begin{eqnarray} h_q(E) &=& \exp_q
\left(-\frac{E}{T}\right) = \frac{2 -
q}{T}\left[1 - (1-q)\frac{E}{T}\right]^{\frac{1}{1-q}} \label{eq:Tsallis}\\
&\stackrel{q \rightarrow 1}{\Longrightarrow}& f(E) =
\frac{1}{T}\exp \left(-\frac{E}{T}\right). \label{eq:BG}
\end{eqnarray}
It is usually argued that, for the $q > 1$ case, $q - 1$ is a
measure of intrinsic fluctuations of the temperature in the system
considered \cite{WW}, whereas  $q < 1$ is usually attributed to
some specific correlations limiting the available phase space
\cite{Kodama} or to the possible fractality of the allowed phase
space \cite{fractal} (other possible interpretations were
considered in \cite{todos})\footnote{One must admit at this point
that this approach is subjected to a rather hot debate of whether
it is consistent with the equilibrium thermodynamics or else it is
only a handy way to a phenomenological description of some
intrinsic fluctuations in the system \cite{debate}. It should be
therefore noticed that it was demonstrated on general grounds
\cite{M} that fluctuation phenomena can be incorporated into a
traditional presentation of a thermodynamic. The Tsallis
distribution (\ref{eq:Tsallis}) belongs to the class of general
admissible distributions which satisfy thermodynamical consistency
conditions and which are therefore a natural extension of the
usual BG canonical distribution (\ref{eq:BG}).}.

For our further considerations of importance are recent
applications of nonextensive statistics in description of nuclear
\cite{Pereira} and quarkonic matter \cite{LPQ,JG1,JG2}, the later
of which we shall continue here. In \cite{Pereira}, the
$q$-version of the Walecka many-body field theory \cite{SW} has
been investigated. It was shown there that $q$-statistics results
in the enhancement of the scalar and vector meson fields in
nuclear matter, in diminishing of the nucleon effective mass and
in hardening of the nuclear equation of state (only the $q > 1$
case was considered there). In \cite{LPQ} the relativistic
equation of state of hadronic matter and a quark-gluon plasma at
finite temperature and baryon density was investigated in the
framework of nonextensive statistical mechanics. In our work
\cite{JG1} we investigated a nonextensive version of another mean
field theory, namely the QCD-based Nambu - Jona-Lasinio (NJL)
model of a many-body field theory describing the behavior of
strongly interacting matter presented recently in \cite{Sousa}.
This time, unlike in \cite{Pereira}, we used the quark rather than
the hadronic degrees of freedom and, because of this, we had to
consider both the $q > 1$ and $q < 1$ cases. This $q$-NJL model
allowed us to discuss the $q$-dependence of the chiral phase
transition in dense quark matter, in particular the quark
condensates and the effective quark masses and their influence on
the masses of $\pi$ and $\sigma$ mesons and on the spinodal
decomposition (cf., \cite{JG1} for details). These results helped
us proceed further and consider critical phenomena in strongly
interaction matter using $q$-statistics (these phenomena are of
interest nowadays, cf., for example, \cite{HK,SFR}, but were so
far not investigated in non-equilibrium environment provided by
$q$-statistics). In particular, we shall now concentrate on the
influence of dynamical factors causing nonextensivity and
represented by the parameter $q$ in the vicinity of the {\it
critical end point} (CEP).

\section{Basic elements of the $q$-NJL model}
\label{sec:II}

First we present the basic elements of the $q$-NJL model
introduced in \cite{JG1} (to which we refer for more details).

\subsection{The usual NJL model}
\label{IIa}

We start with the usual QCD based NJL model based on BG statistics
discussed in \cite{Sousa}. It is the standard $SU(3)$ NJL model
with $U(1)_A$ symmetry described in, with the usual Lagrangian of
the NJL model used in a form suitable for the bosonization
procedure (with four quarks interactions only), from which we
obtain  the gap equations for the constituent quark masses $M_i$:
\begin{eqnarray}
 M_i = m_i - 2g_{_S} \big <\bar{q_i}q_i \big > -2g_{_D}\big
 <\bar{q_j}q_j\big > \big <\bar{q_k}q_k \big >\,,\label{gap}
 \end{eqnarray}
with cyclic permutation of $i,j,k =u,d,s$ and with the quark
condensates given by $\big <\bar{q}_i q_i \big > = -i \mbox{Tr}[
S_i(p)]$ ($S_i(p)$ is the quark Green function); $m_i$ denotes the
current mass of quark of flavor $i$. We consider a system of
volume $V$, temperature $T$ and the $i^{th}$ quark chemical
potential $\mu_i$ characterized by the baryonic thermodynamic
potential of the grand canonical ensemble (with quark density
equal to $\rho_i = N_i/V$, the baryonic chemical potential $\mu_B=
\frac{1}{3} (\mu_u+\mu_d+\mu_s)$ and the baryonic matter density
as $\rho_B = \frac{1}{3}(\rho_u+\rho_d+\rho_s)$),
\begin{equation}
\Omega (T, V, \mu_i )= E- TS - \sum_{i=u,d,s} \mu _{i} N_{i} .
\label{tpot}
\end{equation}
The internal energy, $E$, the entropy, $S$, and the particle
number, $N_i$, are given by \cite{Sousa} (here $E_i = \sqrt{M_i^2
+ p^2}$):
\begin{eqnarray}
E &=&- \frac{ N_c}{\pi^2} V\sum_{i=u,d,s}\left[
   \int p^2 dp  \frac{p^2 + m_{i} M_{i}}{E_{i}}
   (1 - n_{i}- \bar{n}_{i}) \right] - \nonumber\\
   && - g_{S} V \sum_{i=u,d,s}\, \left(\big <
\bar{q}_{i}q_{i}\big > \right)^{2}
   - 2 g_{D}V \big < \bar{u}u\big > \big < \bar{d}d\big > \big <
\bar{s}s\big > , \label{eq:energy} \\
 S &=& -\frac{ N_c}{\pi^2} V \sum_{i=u,d,s}\int p^2 dp \cdot
 \tilde{S}, \label{eq:entropy}\\
 && {\rm where}\quad \tilde{S} =  \bigl[ n_{i} \ln n_{i} + (1-n_{i})\ln (1-n_{i})
   \bigr] +\nonumber\\
   &&~~~~~~~~~~~~~~~~~~~~+ \bigl[ n_{i}\rightarrow 1 - \bar n_{i} \bigr],\nonumber\\
N_i &=& \frac{ N_c}{\pi^2} V \int p^2 dp
  \left( n_{i}-\bar n_{i} \right) \label{number}.
\end{eqnarray}
The quark and antiquark occupation numbers are, respectively,
\begin{eqnarray}
n_{i} &=& \frac{1}{\left\{ \exp\left[\beta \left(E_{i} -
\mu_{i}\right)\right] - 1\right\}}, \label{eq:n}\\
\bar n_{i} &=& \frac{1}{\left\{\exp\left[ \left( \beta(E_{i} +
\mu_{i} \right)\right] + 1\right\}}, \label{eq:barn}
\end{eqnarray}
and with them one calculates values of the quark condensates
present in Eq. (\ref{gap}),
\begin{eqnarray}
\big <\bar{q}_i  q_i \big> = - \frac{ N_c}{\pi^2} \,
\sum_{i=u,d,s}\left[ \int \frac{p^2M_i}{E_i} (1\,-\,n_{i}-\bar
n_{i})\right]dp .\label{gap1}
\end{eqnarray}
Eqs. (\ref{gap}) and (\ref{gap1}) form a self consistent set of
equations from which one gets the effective quark masses $M_i$ and
values of the corresponding quark condensates.

The values of the pressure, $P$, and the energy density,
$\epsilon$, are defined as:
\begin{eqnarray}
 P(\mu_i, T) &=& - \frac{\Omega(\mu_i, T)}{V},\qquad
   \epsilon(\mu_i, T) = \frac{E(\mu_i, T)}{V} \label{eq:p}\\
   ~~{\rm with}~~&& P(0,0) = \epsilon(0,0)=0.\nonumber
\end{eqnarray}

\subsection{The $q$ extension of the NJL model - the $q$-NJL}
\label{IIb}

The $q$-statistics is introduced by using the $q$-form of quantum
distributions for fermions $(+1)$ and bosons $(-1)$ in Eqs.
(\ref{eq:n}) and (\ref{eq:barn}). This is done following a
prescription provided in \cite{TPM}, namely by replacing $n$ and
$\bar{n}$ by
\begin{eqnarray}
n_{qi} &=& \frac{1}{\tilde{e}_q(\beta(E_{qi} - \mu_i))\pm
1},\label{nq} \label{TPM}
\end{eqnarray}
(the important point to notice is that one encounters here $E_{qi}
= \sqrt{M^2_{qi} + p^2}$, i.e., that because of $M_{qi}$ also
energy is now a $q$-dependent quantity). Denoting $x = \beta(E
-\mu)$ one has that for $q > 1$
\begin{eqnarray}
\tilde{e}_q(x) &=& \left\{
\begin{array}{l}
~[1+(q-1)x]^{\frac{1}{q-1}}\quad {\rm if}\quad x > 0 \\
\\
~[1+(1-q)x]^{\frac{1}{1-q}}\quad {\rm if}\quad x\leq 0 \\
\end{array}
\right. , \label{qgt1}
\end{eqnarray}
whereas for $ q < 1$
\begin{eqnarray}
\tilde{e}_q(x) &=& \left\{
\begin{array}{l}
~[1+(q-1)x]^{\frac{1}{q-1}}\quad {\rm if}\quad x \leq 0 \\
\\
~[1+(1-q)x]^{\frac{1}{1-q}}\quad {\rm if}\quad x > 0 \\
\end{array}
\right. .\label{qst1}
\end{eqnarray}
This is because only then can one consistently treat on the same
footing quarks and antiquarks (and for all values of $x$). This
should show the particle-hole symmetry observed in the $q$-Fermi
distribution in plasma containing both particles and
antiparticles, namely that
\begin{equation}
n_q(E,\beta,\mu,q) = 1 - n_{2-q}(- E,\beta,-\mu).
\label{pap_symmetry}
\end{equation}
This means, therefore, that in a system containing both particles
and antiparticles (as in our case) both $q$ and $2 - q$ occur (or,
when expressed by a single $q$ only, one can encounter both $q
> 1$ and $q <1$ at the same time). These dual possibilities warn
us that not only $q > 1$ but also $ q < 1$ (or $(2 - q) > 1$ have
physical meaning in the systems we are considering. This
differentiates  our $q$-NJL model from the $q$-version of the
model presented in \cite{Pereira}. Notice that for $q\rightarrow
1$ one recovers the standard FD distribution, $n(\mu,T)$.
Actually, it is important to realize that for $T\rightarrow0$ one
always gets $n_q(\mu,T)\rightarrow n(\mu,T)$, irrespectively of
the value of $q$ \cite{Pereira}. This means that we can expect any
nonextensive signature only for high enough temperatures (how high
depends on circumstances and on the kind of observable considered,
for illustration of this point see results presented in our paper
\cite{JG1} and Fig. \ref{Figure2} below).

Our $q$-NJL model is then obtained by replacing the formulas of
Section \ref{IIa} with their $q$-counterparts in what concerns the
form of the FD distributions.  Additionally, when calculating
energies and condensates we follow \cite{AL,LPQ} and use the
$q$-versions quark condensates, replacing Eqs. (\ref{gap1}),
(\ref{gap}) and (\ref{eq:energy}) by their $q$-forms:
\begin{equation}
 \big <\bar{q}_i
q_i \big>_q  =  - \frac{ N_c}{\pi^2}  \sum_{i=u,d,s}\left[ \int
\frac{p^2M_{qi}}{E_{qi}} (1\,-\,n^q_{qi}- \bar{n}^q_{qi})\right]dp
,\label{q_gap1}
\end{equation}
\begin{equation}
M_{qi} = m_i - 2g_{_S} \big <\bar{q_i}q_i \big >_q -2g_{_D}\big
 <\bar{q_j}q_j\big >_q \big <\bar{q_k}q_k \big >_q\, ,\label{q_gap}
\end{equation}
\begin{eqnarray}
E_q &=& - \frac{ N_c}{\pi^2} V\!\!\!\sum_{i=u,d,s}\left[
   \int p^2 dp  \frac{p^2 + m_{i} M_{qi}}{E_{qi}}
   (1 - n^q_{qi}- \bar{n}^q_{qi}) \right] - \nonumber\\
   && - g_{S} V \sum_{i=u,d,s} \left(\big <
\bar{q}_{i}q_{i}\big >_q \right)^{2}
   - 2 g_{D}V \big < \bar{u}u\big >_q \big < \bar{d}d\big >_q \big <
\bar{s}s\big >_q . \label{q_energy}
\end{eqnarray}
On the other hand, again following \cite{AL,LPQ}, densities which
are given by the the $q$-version of Eq. (\ref{number}) are
calculated with $n_q$'s (not with $n_q^q$, as in (\ref{q_energy})
and in (\ref{q_gap1})). The pressure for given $q$ is calculated
using the above $E_q$ and the $q$-entropy version of Eq.
(\ref{eq:entropy}) with (cf. \cite{TPM})
\begin{eqnarray}
\tilde{S}_q &=& \left[ n^q_{qi} \ln_q n_{qi} + (1-n_{qi})^q\ln_q
(1-n_{qi}) \right] + \nonumber\\
&&+ \left\{ n_{qi}\rightarrow 1\! -\! \bar n_{qi} \right\}.
\label{q entropy}
\end{eqnarray}
Eq. (\ref{q_gap1}) together with the $q$-version of the gap
equation, Eq. (\ref{q_gap}), are the basic equations from which
one deduces all results presented here.

\section{Results}
\label{sec:Results}

\begin{figure}[h]
  \begin{center}
  \resizebox{0.45\textwidth}{!}{\includegraphics{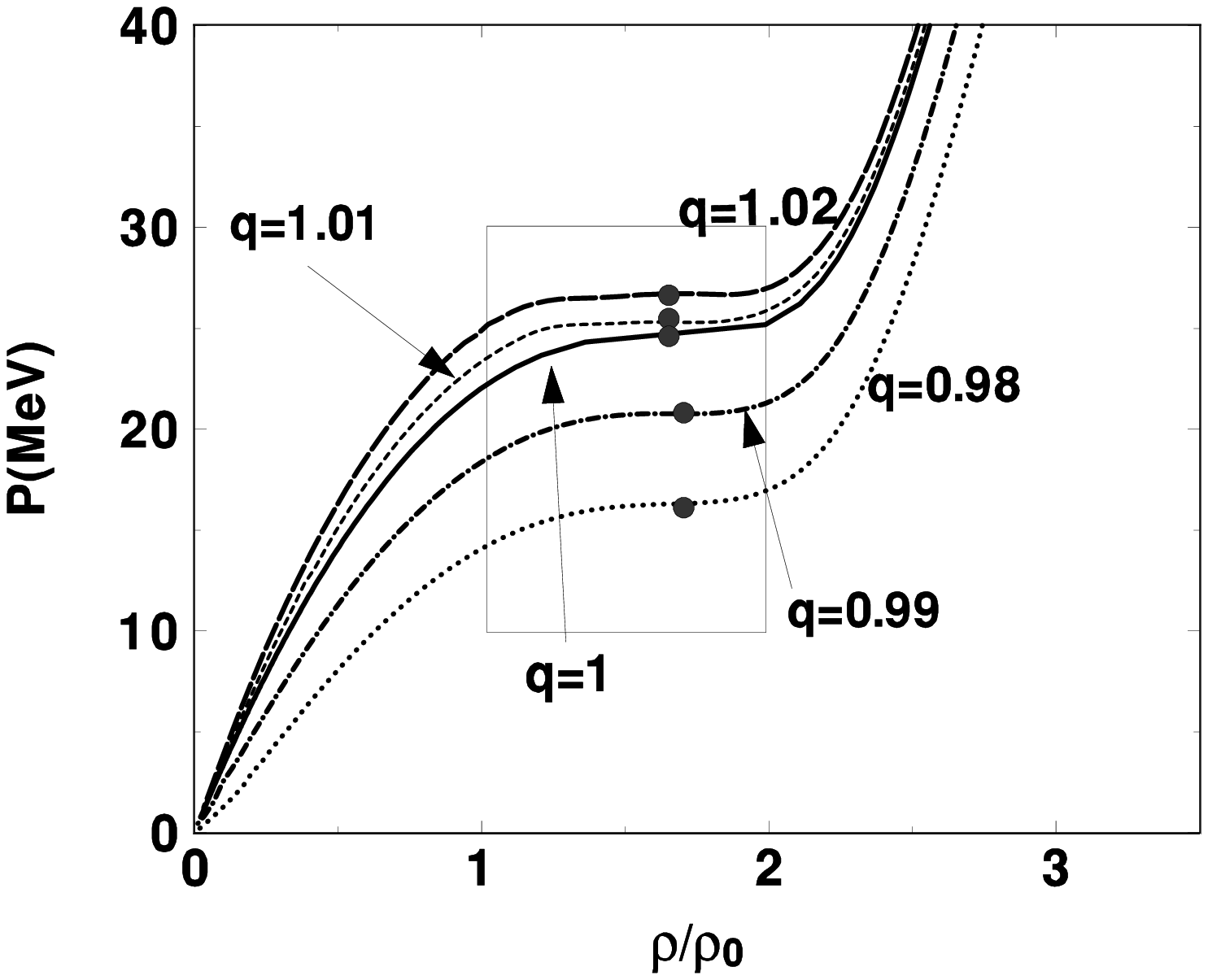}}\\
  \resizebox{0.45\textwidth}{!}{\includegraphics{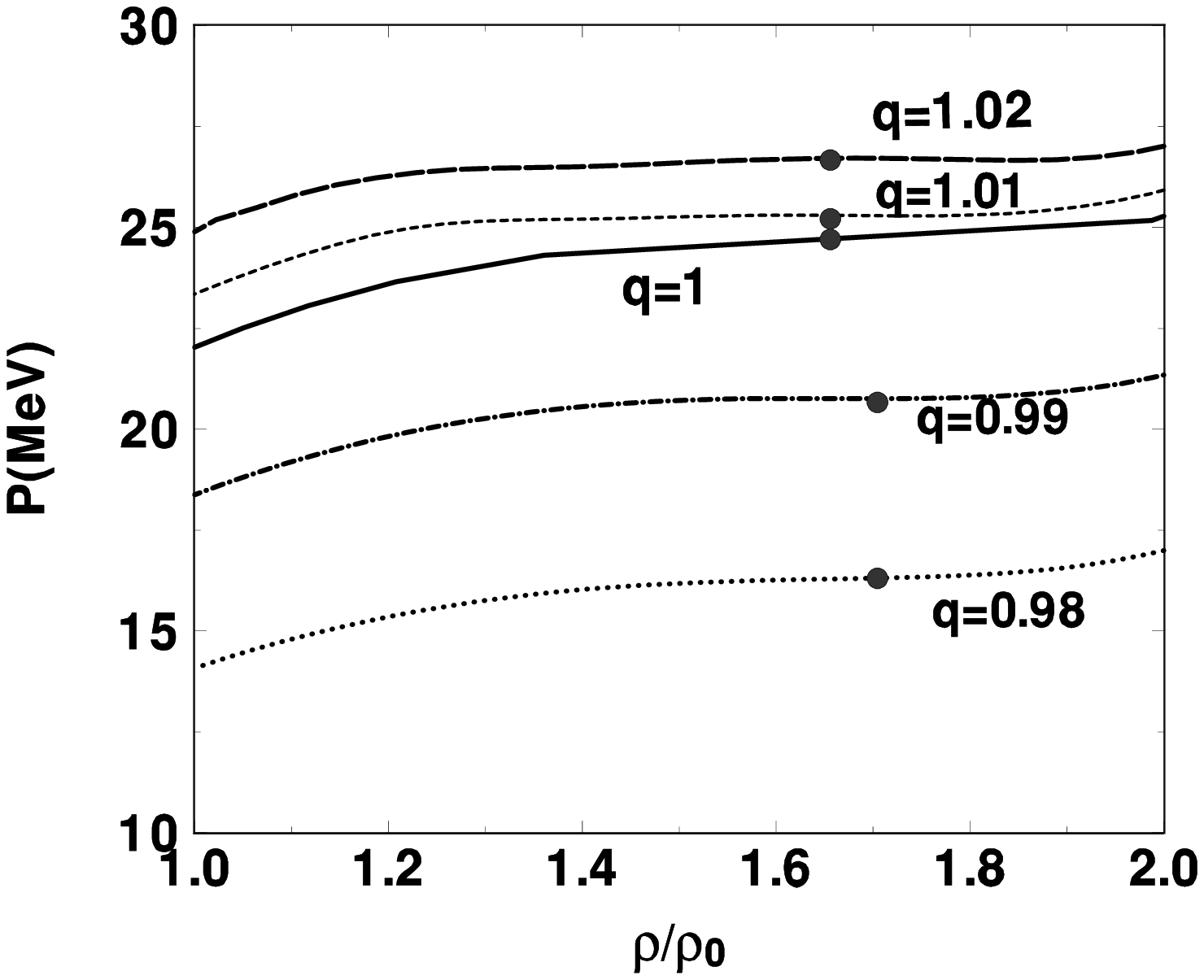}}
  \end{center}
  \caption{The pressure at critical temperature $T_{cr}$ as a
            function of compression $\rho/\rho_0$ calculated
            for different values of the nonextensivity parameter
            $q$ (the area marked at the upper panel is shown in
            detail in the lower panel).
            The dots indicate positions of the inflection
            points for which first derivative of pressure by
            compression vanishes. As in \cite{Sousa} for $q = 1$
            the corresponding compression is $\rho/\rho_0 = 1.67$
            (and this leads to $\mu = 318.5$ MeV); it remains the
            same for $q > 1$ considered here (but now $\mu = 321$
            MeV for $q = 1.01$ and $\mu = 326.1$ MeV for $q = 1.02$)
            whereas it is shifted to $\rho/\rho_0 = 1.72$ for
            $ q< 1$ ($\mu = 313$ MeV for $q = 0.99$
            and $\mu = 307.7$ MeV for $q = 0.98$).
            }
  \label{Figure1}
\end{figure}

Before presenting our results concerning nonextensive critical
effects we shortly repeat the previous results (cf.,
\cite{JG1,JG2}). In Fig. \ref{Figure1} we present the typical
pressure at critical temperature $T_{cr}$ obtained in a $q$-NJL
model as a function of compression $\rho/\rho_0$ calculated for
different values of the nonextensivity parameter $q$
(see \cite{JG1} for more details on spinodial decomposition and
chiral symmetry restoration in $q$-NJL model)\footnote{There is
still an ongoing discussion on the meaning of the temperature in
nonextensive systems. However, in our case the small values of the
parameter $q$ deduced from the data allow us to argue that, to
first approximation, $T_q = T$ used here and in \cite{Pereira}. In
high energy physics it is just the hadronizing temperature (and
instead of the state of equilibrium one deals there with some kind
of stationary state). For a thorough discussion of the temperature
of nonextensive systems, see \cite{Abe}.}.  Notice that the effect
is stronger for $ q <1$ and that, essentially, the saddle point
remains at the same value of compression. When one moves away from
the critical temperature, the typical spinodal structure occurs,
which is more pronounced for lower temperatures whereas its
sensitivity to the $q$ parameter gets stronger with increasing
temperature (cf., \cite{JG1}). However, it occurs that, for each
temperature (even for very small one) a $q > 1$ exists for which
there is no more mixed phase and for which the spinodal effect
vanishes. This seems to be a quite natural effect in the scenario
in which $q > 1$ is attributed to the fluctuations of the
temperature in a system considered as proposed in \cite{WW}. On
the contrary, effects like correlations or limitations of the
phase space considered in \cite{Kodama,fractal} work towards an
increase of the $T_{cr}$ and make the spinodal effect more
pronounced.

\begin{figure}[h]
\begin{center}
\resizebox{0.45\textwidth}{!}{\includegraphics{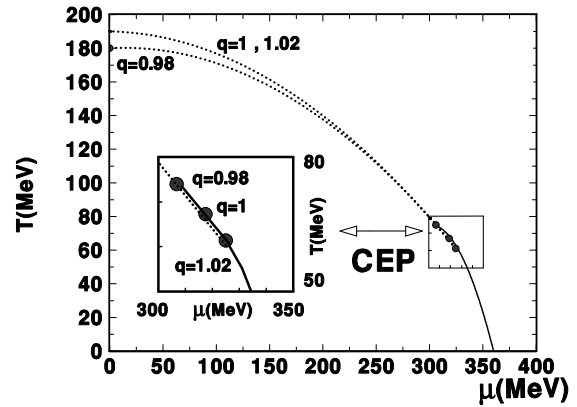}}
\end{center}
\vspace{-0.5cm} \caption{Phase diagram in the $q$-NJL model in $T
- \mu$ plane for values of $q$ considered before:
$q=0.98,~1.0~,~1.02$. Solid and dashed lines denote, respectively,
first order and crossover phase transitions. The results are
presented for three different values of the nonextensivity
parameter $q$ with the vicinity of the ($q$-dependent) critical
end points (CEP) enlarged in the inlet. The crossover phase
transition for $q = 0.98$ and for $\mu \rightarrow 0$ takes place
for a smaller temperature $T$.} \label{Figure2}
\end{figure}

A few remarks are in order here (for more detailed discussion we
refer to \cite{JG1}). Nonextensive dynamics enter the NJL
calculations through the quark (antiquark) number distribution
functions $n_{qi}$ ($\bar{n}_{qi}$). These functions are connected
with the respective quark (antiquarks) spectral functions in the
NJL model. However, deviations from the exponential shape of
$q$-exponents, as defined in Eqs. (\ref{qgt1}) and (\ref{qst1}),
are negligible for values of $q$ close to unity (in our case $0.98
< q < 1.02$). It is also important to notice that Eqs.
(\ref{qgt1}) and (\ref{qst1}) are symmetric for $q \leftrightarrow
1-q$. The differences between $q<1$ and $q>1$ cases observed in
our results are then due to our way of defining the energy
(\ref{q_energy}) and entropy (\ref{q entropy}), which, following
\cite{AL,LPQ}, we do by using $n^q{_{qi}}$ and $\bar{n}^q{_{qi}}$
instead of $n_{qi}$ and $\bar{n}_{qi}$ \footnote{It is worth
notice that in \cite{Pereira}, which considers only the $q>1$ case
and uses number distributions without powers of $q$, the
significant effects were obtained only for much larger values of
the nonextensive parameter $q=1.2$.}. Because now for $q<1$
distributions $n^q{_{qi}}$ and $\bar{n}^q{_{qi}}$ are closer to
unity than $n{_{qi}}$ and $\bar{n}{_{qi}}$, therefore the absolute
values of quark condensates (as given by Eq. (\ref{q_gap1})) begin
to decrease for $q=0.98$ at lower temperature as compared with the
$q=1$ case. Therefore the corresponding energy is larger, which
means that $q < 1$ introduces some residual attractive
correlations which rise the energy and lead to hadronization
occurring at lower temperature. On the other hand, $ q
> 1$ introduces fluctuations which decrease the effective
occupations ($n^q{_{qi}}$ and $\bar{n}^q{_{qi}}$) and the energy,
and smears out the chiral phase transition. In Fig. \ref{Figure2}
we present our phase diagram in the $\mu-T$ plane for different
nonextensivity parameters considered here with positions of the
corresponding critical end points (CEP) for different values of
$q$ clearly indicated.  The overlap of curves observed in Fig.
\ref{Figure2} (inlet) indicates how the critical end point is
smeared to a kind of critical area. This is because fireballs
created in different events can have different values of $q$
(representing, as mentioned before, action of all factors
responsible for the departure of our system from the usual BG
approach - not specified here in detail but, in general, resulting
in specific correlations of quarks or fluctuations of temperature
mentioned before). Therefore when analyzing experimental data one
most probably will encounter such a critical area instead of a
well defined CEP.

\begin{figure*}[t]
  \begin{center}
  \resizebox{0.45\textwidth}{!}{\includegraphics{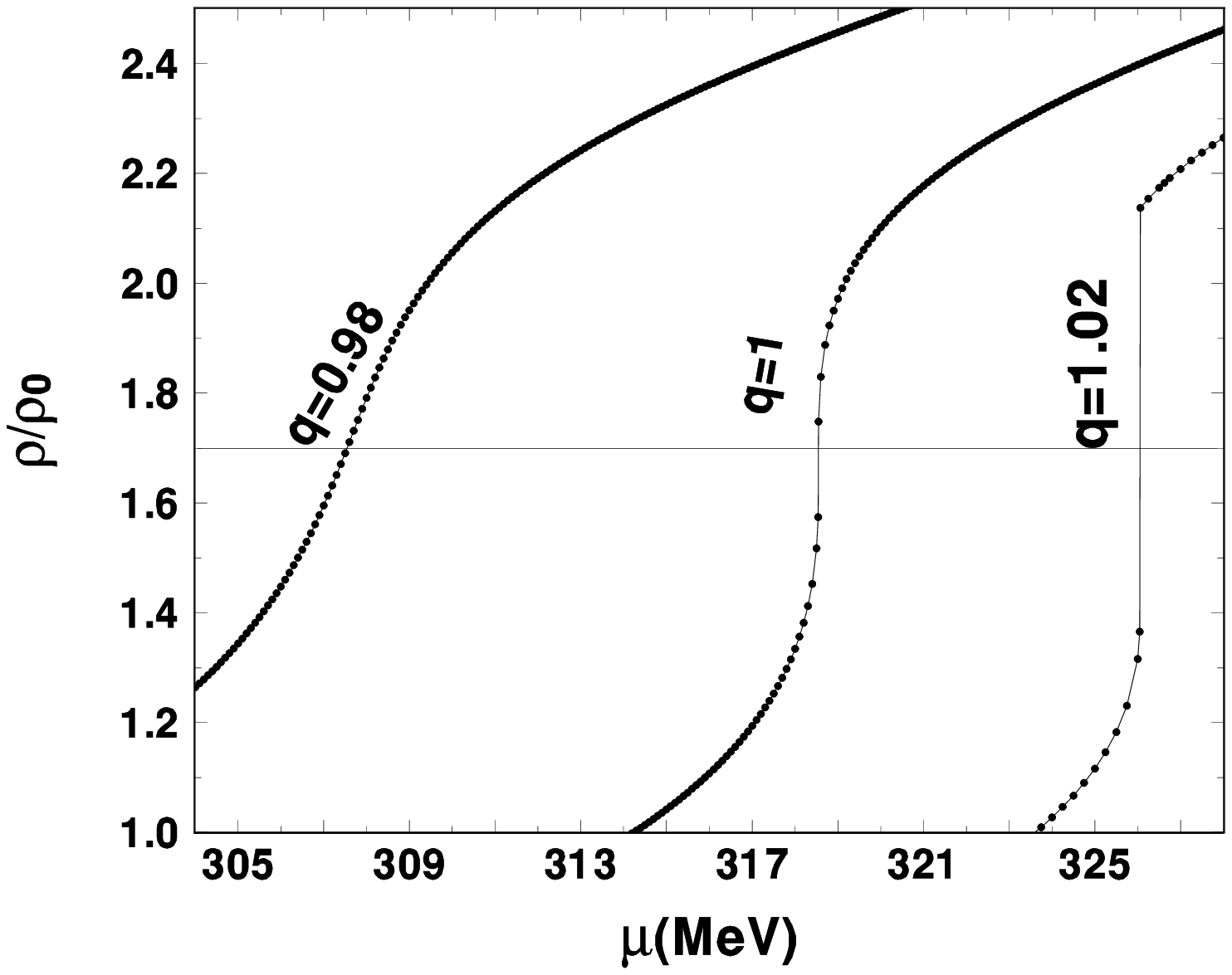}}
  \resizebox{0.45\textwidth}{!}{\includegraphics{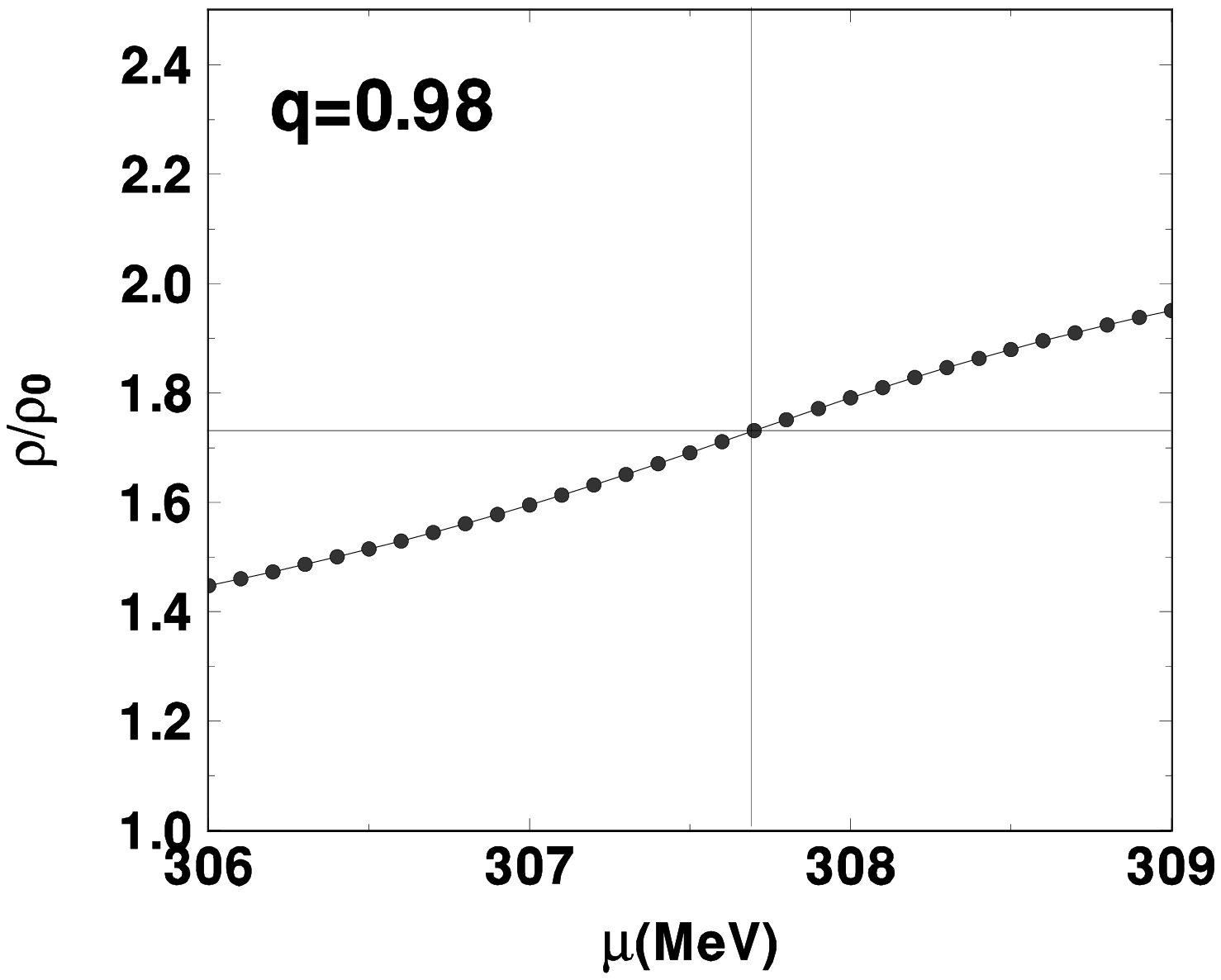}}
  \resizebox{0.45\textwidth}{!}{\includegraphics{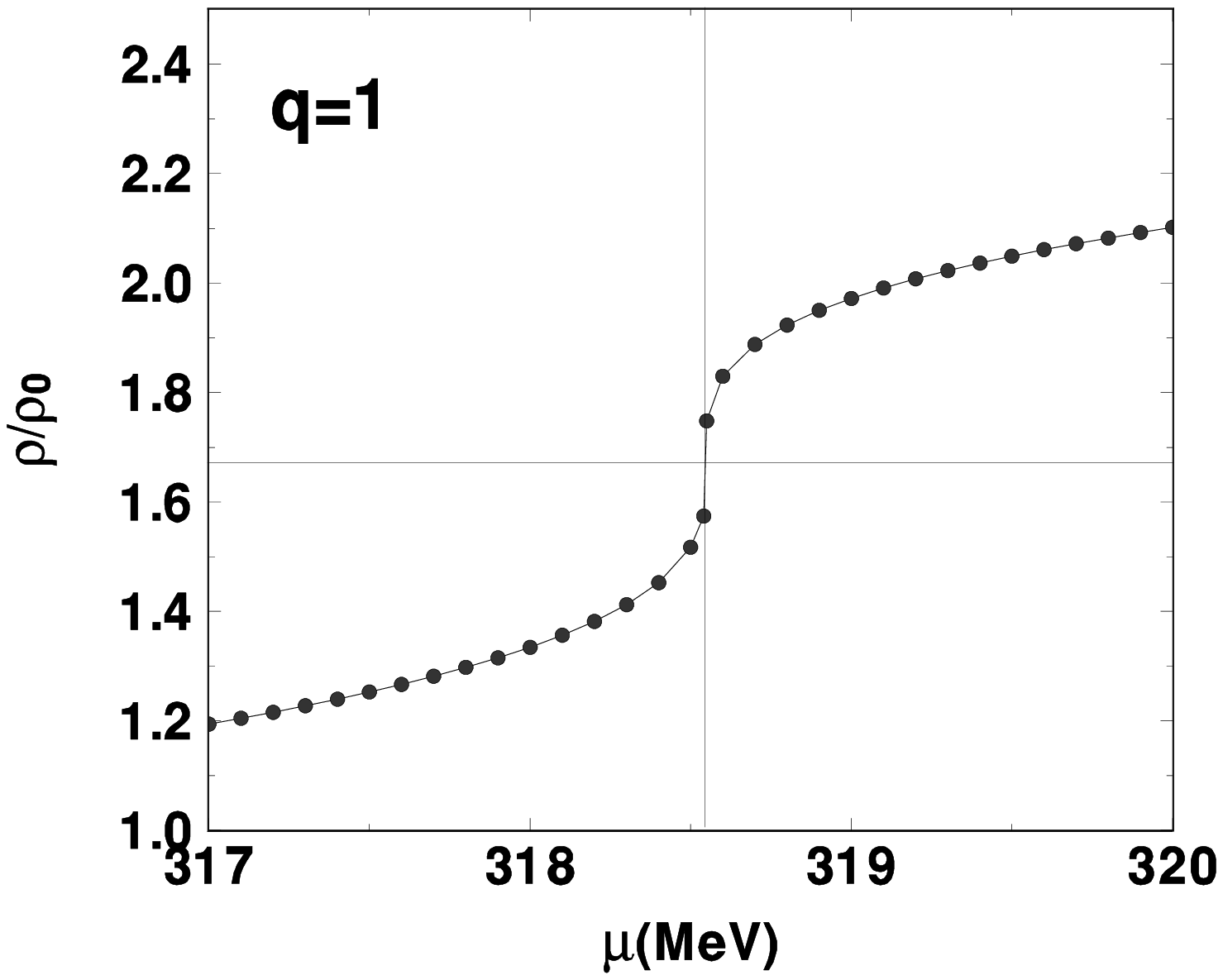}}
  \resizebox{0.45\textwidth}{!}{\includegraphics{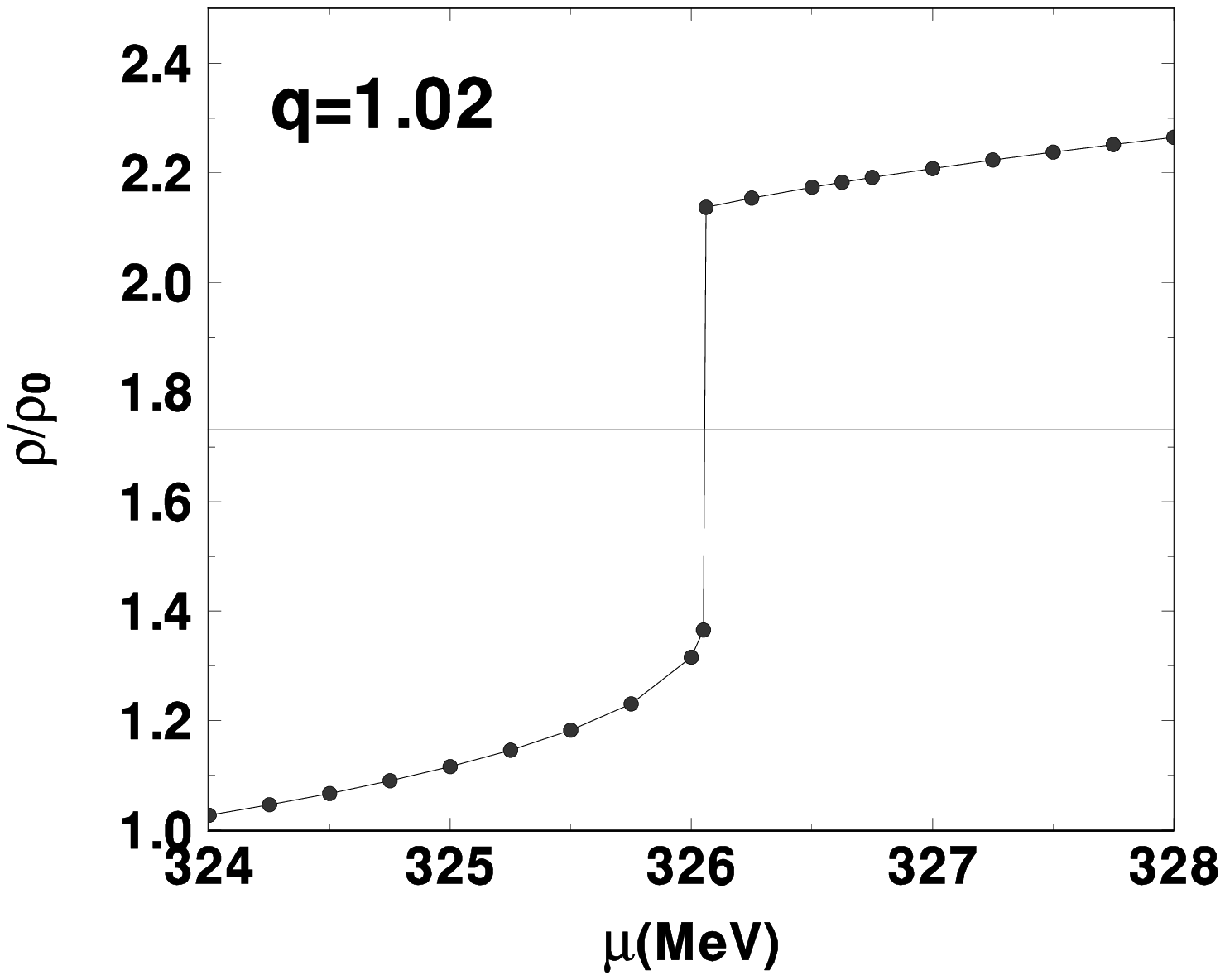}}
  \end{center}
  \caption{The baryon compression $\rho/\rho_0$ (calculated in the vicinity of the
   critical values of temperature and density indicated by the corresponding
   dotted lines) as function of the chemical potential $\mu$ for different values of the
   nonextensivity parameter, $q =0.98, 1.00, 1.02$. The summary presented in the
   top-left panel is detailed in the three consecutive panels.
           }
  \label{Figure3}
\end{figure*}

\begin{figure}[h]
  \begin{center}
\resizebox{0.45\textwidth}{!}{\includegraphics{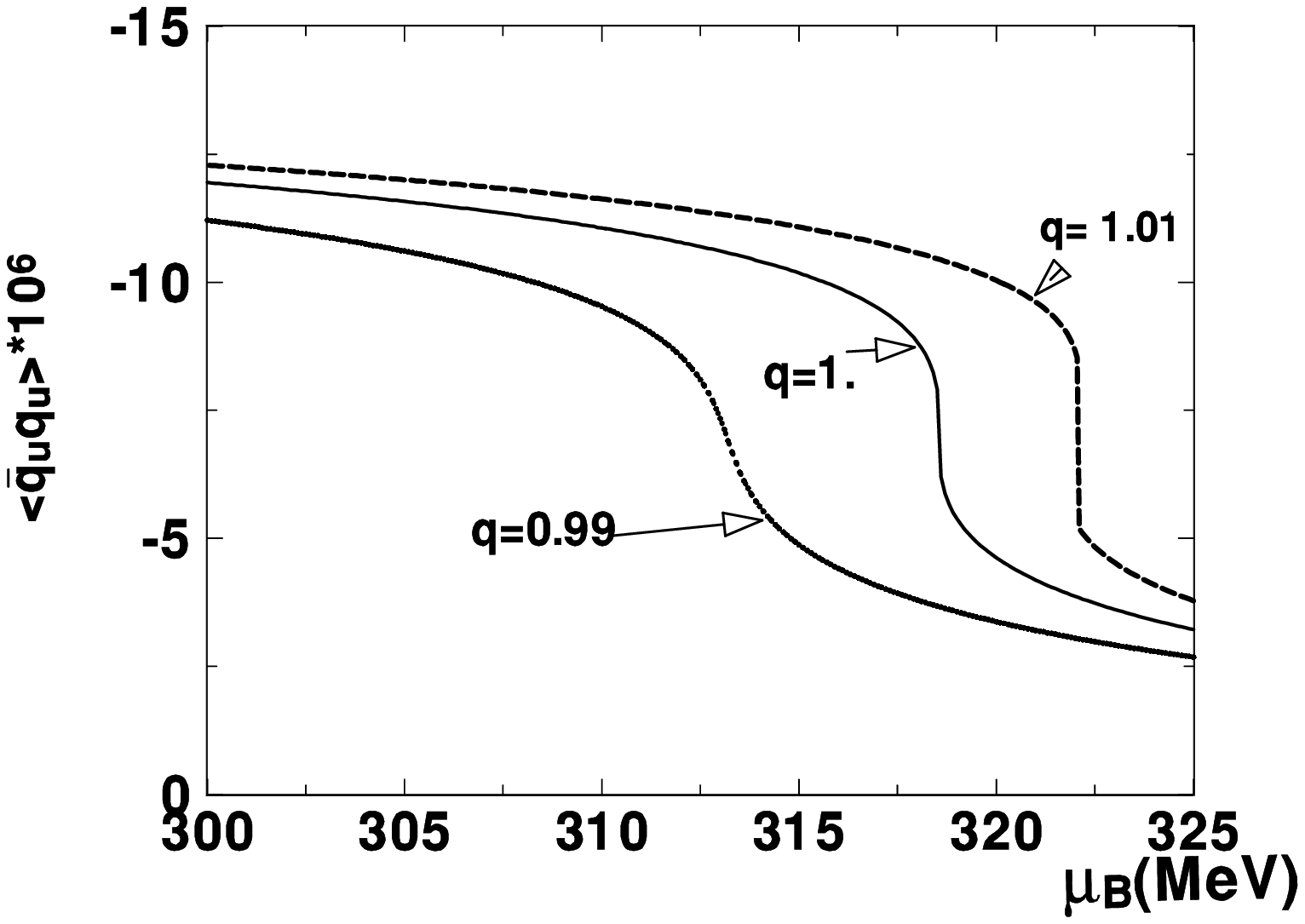}}
\resizebox{0.45\textwidth}{!}{\includegraphics{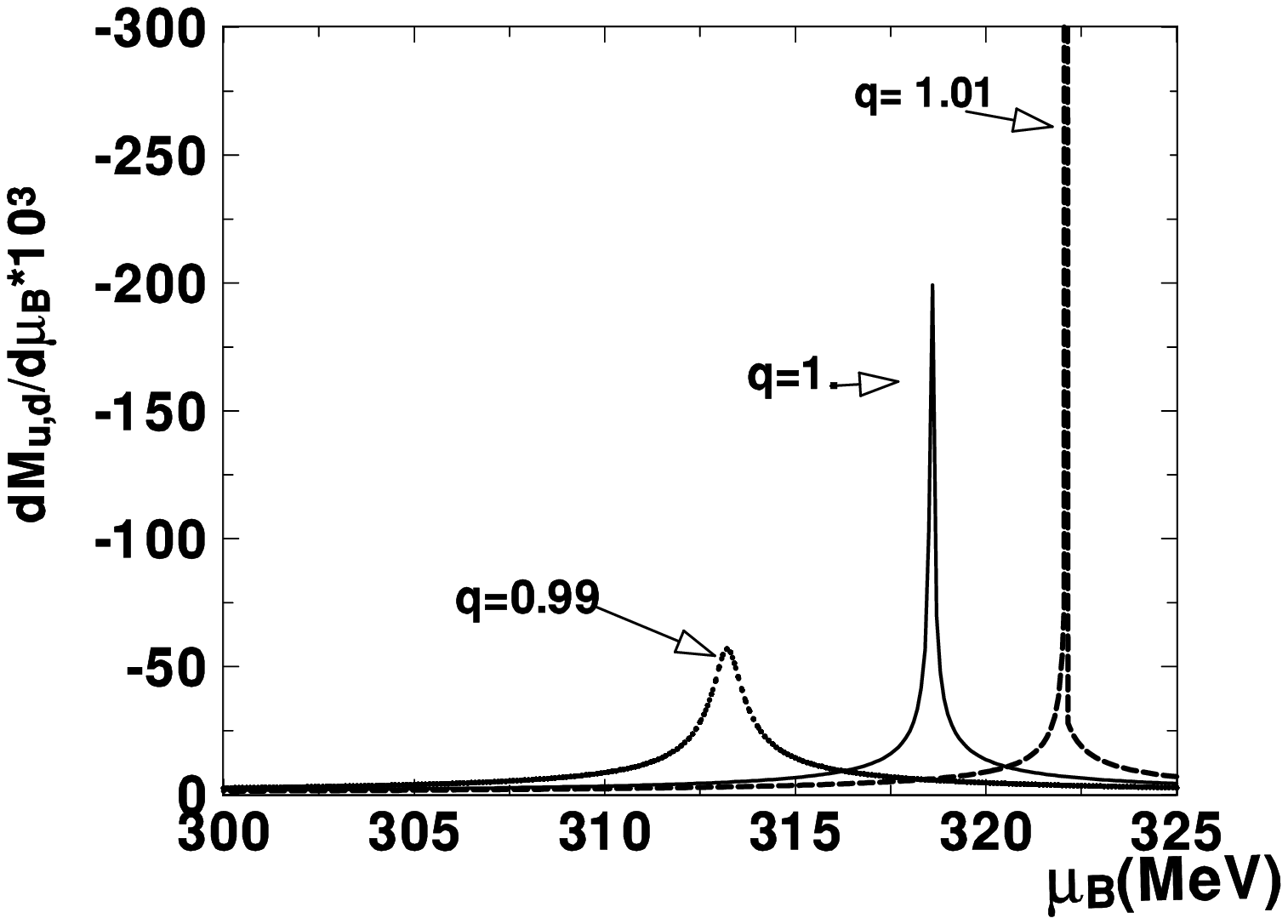}}
  \end{center}
  \caption{Upper panel: the chemical potential ($\mu_B$)  dependence
           of the light quarks condensate in the vicinity of the
           critical region calculated according to Eq. (\ref{q_gap1})
           for different values of the nonextensivity parameter $q$:
           $q = 0.99$, $1.0$ and $1.01$.
           Bottom panel:  the $\mu_B$ dependence of the
           chemical potential derivative of the light quark mass
           $M_{qu}$ calculated according to Eq. (\ref{q_gap})
           in the critical region for the same values of $q$ as above.}.
  \label{Figure4}
\end{figure}

The role of all these factors is shown in more detail in Fig.
\ref{Figure3} which shows the baryon compression $\rho/\rho_0$
(calculated in the vicinity of the critical values of temperature
and density indicated by the corresponding dotted lines) as the
function of the chemical potential $\mu$ for different values of
the nonextensivity parameter, $q =0.98, 1.00, 1.02$. Notice the
remarkable difference of the density derivative at the critical
point: from the smooth transition through the critical point for
$q<1$ to a big jump in density for critical value of chemical
potential for $q>1$. It reflects the infinite values of the baryon
number susceptibility, $\chi_B$:
\begin{equation}
\chi_B = \sum_{i=u,d,s} \left(
\frac{\partial\rho_i}{\partial\mu_B}\right)_T = - \sum_{i=u,d,s}
\left( \frac{\partial^2\Omega}{\partial^2\mu_B}\right)_T.
\label{eq:sus}
\end{equation}
The transition between confined and deconfined phases and/or
chiral phase transition \cite{HK} can be seen  by measuring, event
by event, the difference in the magnitude of local fluctuation of
the net baryon number in a heavy ion collision \cite{Hatta}. They
are initiated and driven mainly by the quark number fluctuation,
described here by $\chi_B$, and can survive through the freezout
\cite{Hatta}. Consequently, our q-NJL model allows us to make the
fine tuning for the magnitude of  baryon number fluctuations
(measured, for example, by the charge fluctuations of protons) and
to find the value of the parameter $q$ characteristic for this
system. However, it does not allow us to differentiate between
possible dynamical mechanisms of baryon fluctuation. We close by
noticing that using $q$ dependent $\chi_B$ leads to $q$-dependent
parameter $\epsilon$ of the critical exponents which describe the
behavior of baryon number susceptibilities near the critical point
\cite{Ikeda}. Whereas in the mean field universality class one has
$\epsilon=\epsilon'=2/3$, our preliminary results using the
$q$-NJL model show a smaller value of this parameter for $q>1$,
($\epsilon \sim 0.6$ for q=1.02) and greater for $q<1$
($\epsilon\sim0.8$ for q=0.98). It would be interesting to deduce
the corresponding values of $q$ from different  models and compare
them with results on a lattice which, by definition, should
correspond to $q=1$ (it should be mentioned at this point that
there are already attempts to apply Monte Carlo methods,
simulating lattice gauge field dynamics as based on non-extensive
rather than extensive thermodynamics, see \cite{Birolat} and
references therein).

\begin{figure}[h]
  \begin{center}
\resizebox{0.45\textwidth}{!}{\includegraphics{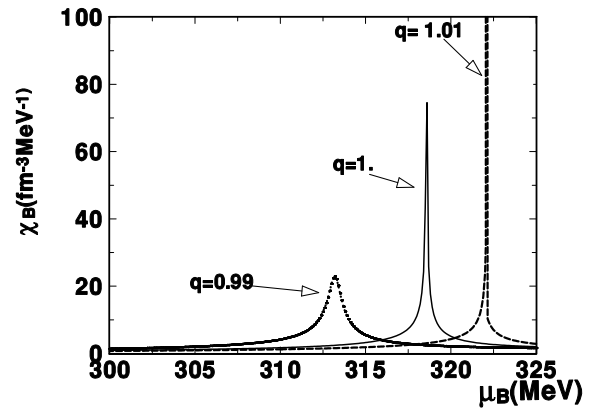}}
  \end{center}
  \caption{The $\mu_B$ dependence of the baryon number susceptibility,
  $\chi_B$, in the vicinity of the critical region, calculated according
  to Eq. (\ref{chi}) for  different nonextensivity nonextensive
  parameters,   $q = 0.99$, $1.0$ and $1.01$. Notice that it is essentially identical
  with results presented in the bottom panel of Fig. \ref{Figure4}.}
  \label{Figure5}
\end{figure}

In order to further investigate the $q$ dependence of $\chi_B$ let
us rewrite Eq. (\ref{eq:sus}) in the following form (recall that
$\rho = N_q/V$ and $N_q$ is $q$-version of Eq. (\ref{number})):
\begin{eqnarray}
\chi_B = \frac{ 1}{\pi^2} \sum_{i=u,d,s} \int p^2 dp \left(
\frac{\partial n_{qi}}{\partial\mu_B} - \frac{\partial\bar
n_{qi}}{\partial\mu_B} \right)_T. \label{chi}
\end{eqnarray}
The $q$-versions of occupation numbers, $n_{qi}$ and
$\bar{n}_{qi}$, are taken from Eq. (\ref{TPM}). The $q$-version of
energies there depend on masses $M_{qi}$, which are given by gap
equation (\ref{q_gap}) in a quite involved way. Therefore, the
$q$-dependence enters here in two ways: by rather straightforward
replacement of $\exp(...)$ by the respective $\exp(...)$ in Eq.
(\ref{TPM}) and by quite involved $q$-dependence of $M_{qi}$ given
by the gap equation (\ref{q_gap}). Therefore,
\begin{eqnarray}
\chi_B(\mu_B,T) &=& \frac{1}{\pi^2 T} \cdot \left[
\chi\left(\mu_B\right)+\bar{\chi}\left(\mu_B\right)\right]
\label{chi1}
\end{eqnarray}
with
\begin{eqnarray}
{\chi}\left(\mu_B\right)\!\! &=&\!\! \sum_{i=u,d,s}\!\! \left[\int
\! p^2 dp n_{qi}^2 \left(\frac{1 -
n_{qi}}{n_{qi}}\right)^{f(q)}\left(1- \frac{M_{qi}}{E_{qi}}
\frac{\partial M_{qi}}{\partial\mu_B}\right)\right], \nonumber\\
\bar{\chi}\left(\mu_B\right)\!\! &=&\!\! \sum_{i=u,d,s}\!\!
   \left[\int\! p^2 dp \bar{n}_{qi}^2 \left(\frac{1 - \bar{n}_{qi}}
   {\bar{n}_{qi}}\right)^{f(q)}
   \left(1+\frac{M_{qi}}{E_{qi}}\frac{\partial
M_{qi}}{\partial\mu_B}\right)\right] ,   \nonumber
\end{eqnarray}
where
\begin{eqnarray}
f(q)&=&(2-q) \qquad {\rm if}\quad (q-1)(E_{qi}-\mu_B) > 0,  \nonumber \\
f(q)&=&q. \qquad \qquad{\rm otherwise}. \nonumber
\end{eqnarray}
Our results are presented in Figs. \ref{Figure4} and
\ref{Figure5}. It turns out that the chiral phase transition
investigated here (Fig. \ref{Figure5}) is mainly driven by the
behavior of the light quark mass derivative, see Fig.
\ref{Figure4}, which in turn is determined by the behavior of the
light condensate, cf., Fig. \ref{Figure3}. Thus the dynamic of the
nonextensive effects is generated not so much by the nonextensive
form of occupation numbers in Eq. (\ref{TPM}) but rather by the
main gap equation (\ref{gap}) where both the condensates and the
effective quark masses are present.

\section{\label{sec:IV}Summary}

We have investigated the sensitivity of critical behavior of the
QCD based NJL type of mean theory type presented in \cite{Sousa},
the $q$-NJL model, to the departure from the conditions required
by the application of the BG approach by using the Tsallis version
of nonextensive statistical mechanics \cite{T}. All factors
causing this departure are summarily described by the
nonextensivity parameter $q$, such that $q-1$ quantifies departure
from the BG situation (which is recovered for $q \to 1$).

We have investigated two possible scenarios corresponding to $q
> 1$ and $q < 1$, respectively, which, as mentioned, correspond to
different physical interpretations of the nonextensivity
parameter. For $ q <1$ (usually connected with some specific
correlations \cite{Kodama} or with fractal character of the phase
space \cite{fractal}) we observe a decreasing of pressure, which
reaches negative values for a broad ($q$-dependent) range of
temperatures and increasing of the critical temperature
\footnote{It acts therefore in the same way as including of the
Polyakov loop into the NJL model \cite{PNJL}.}.  The $ q > 1$ case
(usually connected with some specific nonstatistical fluctuations
existing in the system \cite{WW}) we observe a decreasing of the
critical temperature, $T_{crit}$, and therefore in the limit of
large $q$ we do not have a mixed phase but rather a quark gas in
the deconfined phase above the critical line (on the contrary, the
compression at critical temperature does not depend on $q$. As in
\cite{Pereira} the resulting equation of state is stiffer (in the
sense that for a given density we get larger pressure with
increasing $q$). As expected, the effects depend on the
temperature, and tend to vanish when the temperature approaches
zero. Fig. \ref{Figure3} shows that the nonequilibrium statistics
dilutes the border between the crossover and the first order
transition. Finally, Figs. \ref{Figure4} and \ref{Figure5}
demonstrate that the most important $q$-dependence is coming from
the main gap equation (\ref{gap}), where both the condensates and
the effective quark masses are present, rather than from the the
nonextensive form of occupation numbers in Eq. (\ref{TPM}).

We would like to end by stressing that our results could be of
interest for investigations aimed at finding the critical point in
high energy heavy ion collisions \cite{departure} or when studying
particularities of the equation of state (EoS) of compact stars
\cite{NSTARS}. The fact that they depend on the parameter $q$
means that the exact position of such a point or the type or shape
of EoS could be quite different from what is naively expected.

\section*{Acknowledgements}
Partial support of the Ministry of Science and Higher Education
under contract DPN/N97/CERN/2009 for (GW)  and under the Research
Project No. N N202046237 for (JR) is acknowledged.

\end{document}